\documentclass{aastex631}
\usepackage{CJK, todonotes}
\usepackage{gensymb, amsmath, pifont}

\begin{document}
\begin{CJK*}{UTF8}{gbsn}

\title{Disintegration of Long-period Comet C/2019 Y4 (ATLAS). II. Post-Perihelion Remnant Recovery}

\correspondingauthor{Salvatore A. Cordova Quijano}

\author[0009-0008-5998-0185]{Salvatore A. Cordova Quijano}
\affiliation{Center for Space Physics, Boston University, 725 Commonwealth Ave, Boston, MA 02215, USA}
\email[show]{liebe@bu.edu}

\author[0000-0002-4838-7676]{Quanzhi Ye (叶泉志)}
\affiliation{Department of Astronomy, University of Maryland, College Park, MD 20742, USA}
\affiliation{Center for Space Physics, Boston University, 725 Commonwealth Ave, Boston, MA 02215, USA}
\email{qye@umd.edu}

\author[0000-0002-6702-7676]{Michael S. P. Kelley}
\affiliation{Department of Astronomy, University of Maryland, College Park, MD 20742, USA}
\email{mkelley5@umd.edu}



\begin{abstract}
We present an investigation into the fate of disrupting near-Sun comet C/2019 Y4 (ATLAS). Imaging observations with the Lowell Discovery Telescope (LDT), obtained 3--5 months after the reported disruption and the last sighting, constrained the primary component C/2019 Y4-B to an absolute magnitude of $H>20.5$ or a diameter of $D<0.5$~km (assuming a geometric albedo of 0.04) at an outbound heliocentric distance of 2.71~au. A search of shallower data obtained from the Zwicky Transient Facility (ZTF) survey, conducted when components A and B receded from 2.5 to 2.9~au from the Sun, yielded no detections, suggesting that the fragments had either ceased to exist or were not substantially active during this period. The uncertain fate of C/2019 Y4 highlights the challenge in understanding the state of presumably disrupted comets. Our review of six other presumably disrupted long-period comets reveals that negative confirmations were not reported for half of them, leaving their disruption presumptions inconclusive.
\end{abstract}



\section{Introduction} 

In early 2020, the near-Sun comet C/2019 Y4 (ATLAS) garnered scientific and public interest due to its potential for naked-eye visibility, owing to both its intrinsic brightness and small perihelion distance of $q=0.25$ au \citep{IAU2020}. The comet was discovered by the Asteroid Terrestrial-Impact Last Alert System (ATLAS) search program on 2019 December 28. Shortly after the discovery, M. Meyer identified C/2019 Y4 as possessing similar orbital elements as C/1844 Y1 (Great Comet) \citep{IAU2020}, a finding that suggested the possibility of both comets sharing a common progenitor \citep{hui2020, sekanina2023fragmentation}.

In April of 2020, C/2019 Y4 underwent significant fragmentation into four major discernible fragments: A, B, C, and D \citep{IAU2020b}. In Paper I \citep{Ye2021}, we reported the results from a 3-day Hubble Space Telescope (HST) campaign on the comet during its fragmentation. We found that the complex was dominated by fragment B, a cluster of bright, relatively long-lived mini-comets. Fragment A was another bright fragment cluster that showed rapid mass loss during the 3-day campaign. We also found that fragment C was likely a data artifact, and fragment D only lived for about a week into the fragmentation event. The behavior of the fragments suggested that C/2019 Y4 had a nonuniform interior in which the longer-lived fragment B likely represented the stronger, ``drier'' part of it. 

C/2019 Y4 was last reportedly observed on 2020 June 8 by the Solar TErrestrial RElations Observatory-A (STEREO-A) spacecraft, 9 days after its perihelion passage \citep{2020ATel13813....1K}, displaying a lack of nucleus condensation consistent with a completely disrupted comet, after which no detections had been reported as the comet exited the avoidance zone in August. In order to determine the fate of this comet, we conducted dedicated observations with the Lowell Discovery Telescope (LDT) and analyzed data from the Zwicky Transient Facility (ZTF), with the goal of recovering or at least constraining our knowledge of the remnants of C/2019 Y4.

\section{Observations} \label{sec:style}

\subsection{Lowell Discovery Telescope (LDT)}

We observed the ephemeris position of the primary fragment, C/2019 Y4-B, on 2020 August 20 and October 16 using the Large Monolithic Imager \citep[LMI;][]{Massey2021} on LDT. The observing circumstances are tabulated in Table~\ref{tbl:obs}. The images on August 20 were obtained during twilight, hence an SDSS {\it i'} filter was used to suppress the sky background; the images on October 16 were obtained through an ultra broadband {\it VR} filter. LMI has a field-of-view (FOV) of $12\farcm3 \times 12\farcm3$ and a pixel scale of $0\farcs36$ under an on-chip $3\times3$ binning. We astrometrically calibrated each image using Astrometry.net—a FITS calibration Python package \citep{Lang_2010}. This yielded FITS images with World Coordinate System (WCS) meta-data for precise alignment, facilitating direct comparison of count information between images. Images were bias-subtracted and flat-field corrected using {\tt ccdproc} \citep{Craig2015}, and were then calibrated with the RefCat2 catalog \citep{2018ApJ...867..105T} for photometry.

To enable a more sensitive search, frames from both nights were combined into stacked images using the motion of fragment B calculated from JPL orbit solution \#9. The stacked image represented a total of $2\times3=6$~min and $15\times 2=30$~min integrations for the August 20 and October 16 observations, respectively. The October 16 observation includes a total of 15 frames, sufficient for applying an image differencing technique that provides a cleaner image for examination. We first stacked all 15 frames to create a ``reference'' image, and then subtracted each frame with this reference image using the {\tt HOTPANTS} algorithm \citep{2015ascl.soft04004B}. Finally, we combined all the subtracted images into one stacked, star-removed image following the motion of the comet.

\subsection{Zwicky Transient Facility (ZTF)} 

As the LDT/LMI observations are relatively narrow in terms of field-of-view and only cover two epochs, we also used the survey images obtained by the ZTF project which provided shallower but more frequent coverage. This provides the advantages of (1) examining other fragments which are beyond LDT's field-of-view; and (2) looking for potential outbursts that these fragments may exhibit. ZTF combines the 1.2 m Oschin Schmidt telescope at Palomar Observatory with a 47 deg$^2$ camera to conduct repeat survey imaging of the visible sky. With a typical 30 s exposure, ZTF observes an area of 3760 deg$^2$ per hour down to a $5\sigma$ limit of $r=20.7$. Details of the survey can be found in \citet{Bellm2019}, \citet{Graham2019}, and \citet{Masci2019}.

We used {\it ZChecker} \citep{Kelley2019} to extract ZTF images that contain the ephemeris positions of fragments A and B, in 2020 October (see Table~\ref{tbl:obs}). (We did not investigate fragments C or D as they were either data artifacts or had convincingly disrupted.) No post-perihelion observations exist before 2020 October due to low solar elongation of the comet. We also did not include images beyond 2020 October, as the comet's apparent magnitude would have continued to fade with time at a rate of at least 0.5~mag/month. {\it ZChecker} combines frames from each night into nightly stacks following the motion of the comet, thus allowing a more sensitive search of the fragments. Photometric zero-points of the nightly stacks are calculated by the ZTF Science Data System by comparing field stars with the Pan-STARRS Data Release 1 catalog \citep[][\S~3.5]{Masci2019}.

\begin{deluxetable*}{lccccc}
\tablecaption{Summary of LDT and ZTF observations.}
\label{tbl:obs}
\tablehead{
\colhead{Date \& Time (UT)} & \colhead{$r_\mathrm{h}$ (au)} & \colhead{$\varDelta$ (au)} & \colhead{$\alpha$} & \colhead{Telescope} & \colhead{Filters (exposures, limiting mag, image FWHM)}
}
\startdata
2020 Aug 20 11:45--11:51 & 1.84 & 2.43 & $23^\circ$ & LDT & {\it i} ($2\times180$~s, 19.0, $2\farcs6$) \\
2020 Oct 02 11:05--12:10 & 2.51 & 2.59 & $23^\circ$ & ZTF & {\it g} ($1\times30$~s, 19.6, $2\farcs2$), {\it r} ($1\times30$~s, 19.6, $1\farcs9$) \\
2020 Oct 04 11:35 & 2.54 & 2.57 & $23^\circ$ & ZTF & {\it g} ($1\times30$~s, 19.7, $2\farcs1$) \\
2020 Oct 06 11:09--11:46 & 2.57 & 2.57 & $22^\circ$ & ZTF & {\it g} ($1\times30$~s, 19.0, $2\farcs7$), {\it r} ($1\times30$~s, 19.4, $2\farcs0$) \\
2020 Oct 08 11:12--12:00 & 2.60 & 2.57 & $22^\circ$ & ZTF & {\it g} ($1\times30$~s, 19.9, $2\farcs6$), {\it r} ($1\times30$~s, 20.0, $2\farcs1$) \\
2020 Oct 12 11:48--11:49 & 2.66 & 2.57 & $22^\circ$ & ZTF & {\it r} ($2\times30$~s, 20.3, $2\farcs4$) \\
2020 Oct 13 11:29--12:03 & 2.67 & 2.57 & $22^\circ$ & ZTF & {\it g} ($1\times30$~s, 20.6, $2\farcs2$), {\it r} ($1\times30$~s, 20.5, $1\farcs9$) \\
2020 Oct 14 12:31--12:44 & 2.69 & 2.57 & $22^\circ$ & ZTF & {\it r} ($2\times30$~s, 21.0, $1\farcs4$) \\
2020 Oct 15 10:59--12:02 & 2.70 & 2.57 & $22^\circ$ & ZTF & {\it g} ($1\times30$~s, 20.6, $2\farcs3$), {\it r} ($2\times30$~s, 20.7, $1\farcs7$) \\
2020 Oct 16 10:42--11:15 & 2.71 & 2.57 & $23^\circ$ & LDT & {\it VR} ($15\times120$~s, $r=25.5$, $1\farcs7$) \\
2020 Oct 17 11:15 & 2.73 & 2.56 & $21^\circ$ & ZTF & {\it r} ($1\times30$~s, 20.4, $1\farcs9$) \\
2020 Oct 22 10:58--12:35 & 2.80 & 2.56 & $21^\circ$ & ZTF & {\it g} ($3\times30$~s, 21.2, $2\farcs2$), {\it r} ($3\times30$~s, 21.0, $1\farcs7$) \\
2020 Oct 24 11:34 & 2.83 & 2.56 & $21^\circ$ & ZTF & {\it g} ($1\times30$~s, 16.4, $2\farcs0$) \\
2020 Oct 28 10:30--11:30 & 2.88 & 2.55 & $20^\circ$ & ZTF & {\it g} ($1\times30$~s, 19.9, $1\farcs9$) \\
2020 Oct 29 10:59 & 2.89 & 2.56 & $20^\circ$ & ZTF & {\it r} ($1\times30$~s, 19.7, $2\farcs8$) \\
2020 Oct 30 10:07--10:09 & 2.91 & 2.55 & $20^\circ$ & ZTF & {\it r} ($2\times30$~s, 19.8, $2\farcs6$) \\
2020 Oct 31 09:01--12:08 & 2.92 & 2.55 & $19^\circ$ & ZTF & {\it g} ($1\times30$~s, 19.4, $2\farcs9$), {\it r} ($3\times30$~s, 20.0, $2\farcs4$) \\
\enddata
\tablecomments{$r_\mathrm{h}$, $\varDelta$, and $\alpha$ are heliocentric distance, geocentric distance, and phase angle of C/2019 Y4-B, respectively. We note that the distances of components A and B differ by $\sim0.01$~au during the time of the observation due to the propogation of orbital uncertainty. Limiting magnitudes are appropriate to their respective bandpasses unless otherwise noted. Image FWHM is the full-width-half-maximum of the background stars.}
\end{deluxetable*}

\section{Analysis} 

\subsection{Fragment A}

\begin{figure}
    \centering
    \includegraphics[width=0.5\linewidth]{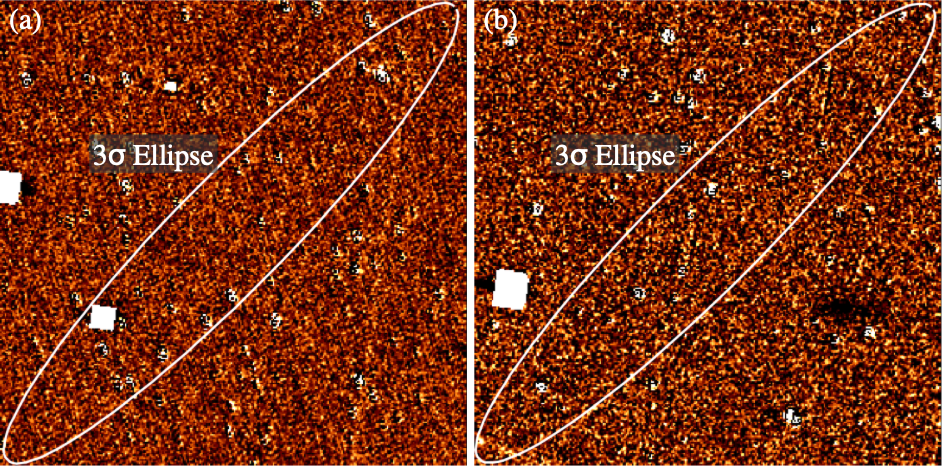}
    \caption{ZTF nightly stacks for C/2019 Y4-A on 2020 October 14 (panel a) and 15 (panel b). Also shown are 3$\sigma$ uncertainty ellipses for fragment A calculated using JPL orbit solution \#7. Each image is $5'\times5'$ in size with up points to celestial north. The white rectangles are masked stars and subtraction residuals.}
    \label{fig:ZTF}
\end{figure}

We visually searched the ZTF nightly stacks for fragment A. Most of the stacks were affected by high sky background and/or a large number of subtraction artifacts. The ones on October 14 and 15 (Figure~\ref{fig:ZTF}) were deeper and were relatively free of artifacts. We concluded that fragment A, if it still existed, was fainter than $r\sim20.5$, equivalent to a diameter of $D\lesssim2$~km. This limit is substantially larger than the $D\lesssim0.7$~km measurement made by the pre-perihelion HST observations as reported in Paper I.

\begin{figure}[h!]
    \centering
    \includegraphics[width=0.5\linewidth]{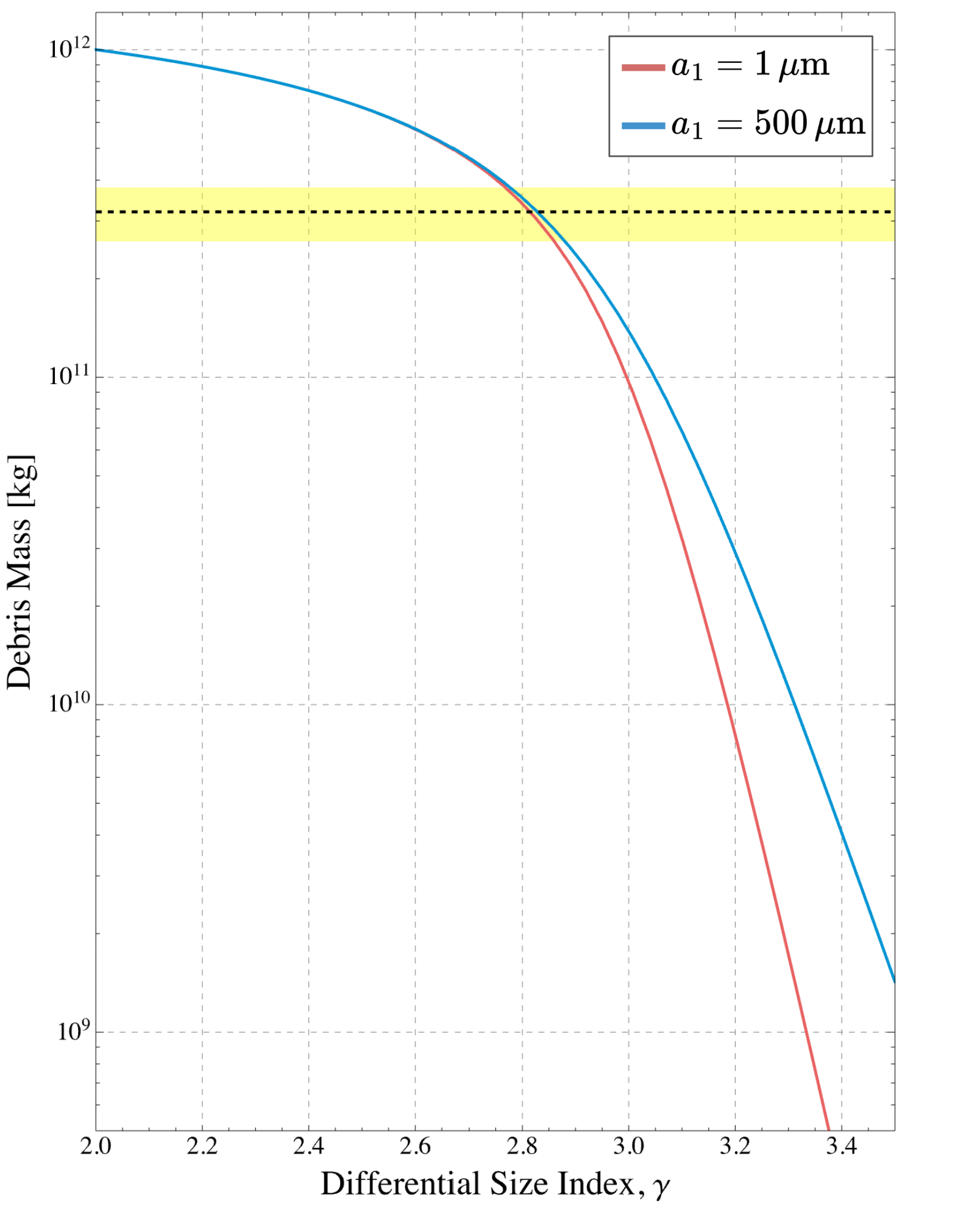}
    \caption{Debris mass of fragment A as a function of differential size index $\gamma$. Two lower bound cases are plotted for $a_1=1 \, \mu\mathrm{m}$ and $a_1=500 \, \mu\mathrm{m}$. The dashed horizontal line, $M_d=(3.2\pm0.6)\times10^{11}$ kg, is derived in Paper I, with the highlighted region representing the constraint's uncertainty.}
    \label{fig: Differential Size Index}
\end{figure}
 
If we assume that the complete disruption occurred not long before the ZTF observation and the remnant was sufficiently compact to be approximated by a point source, the non-detection of fragment A can be used to approximate its differential size distribution index, which characterizes how the total mass of the fragment is distributed among particles of different sizes, indicating whether the debris is dominated by small or large particles. Assuming that the dust size distribution follows a power law, we follow the derivation in \citet{Jewitt_2023} of the combined mass of a comet composed of spherical particles of radii between $a$ and $a+\mathrm{d}a$:
\begin{equation}\label{eqn: Md}
M_d=\frac{4}{3}\rho C\frac{\int_{a_1}^{a_2} a^{3-\gamma}\,\mathrm{d}a}{\int_{a_1}^{a_2} a^{2-\gamma}\, \mathrm{d}a},
\end{equation}
where $\rho$ is the comet's volume mass density (taken to be $\rho\approx500\,\mathrm{kg}/\mathrm{m}^3$), $C$ is the combined geometric cross section of the dust (approximated as $C\approx \pi (D/2)^2\approx3.1\,\mathrm{km}^2$), $\gamma$ is the differential size distribution index, and $a_1$ and $a_2$ are the lower and upper bounds of the radii of the particles respectively. We plot Equation (\ref{eqn: Md}) with upper bound $a_2=D/2=1\,\mathrm{km}$ and two lower bounds $a_1=1\,\mu\mathrm{m}$ and $a_1=500\,\mu\mathrm{m}$ against a mass constraint of $M_d=(3.2\pm0.6)\times10^{11}$ kg derived in Paper I, shown in Figure~\ref{fig: Differential Size Index}. Both lower bound cases intersect the mass constraint at $\gamma=2.8\pm0.1$, indicating that the approximation is insensitive to choice of lower bound. A differential size index of $\gamma<3$ suggests that the mass of fragment A is dominated by large particles.

\subsection{Fragment B}
\subsubsection{LDT}

\begin{figure}
    \centering
    \includegraphics[width=1\linewidth]{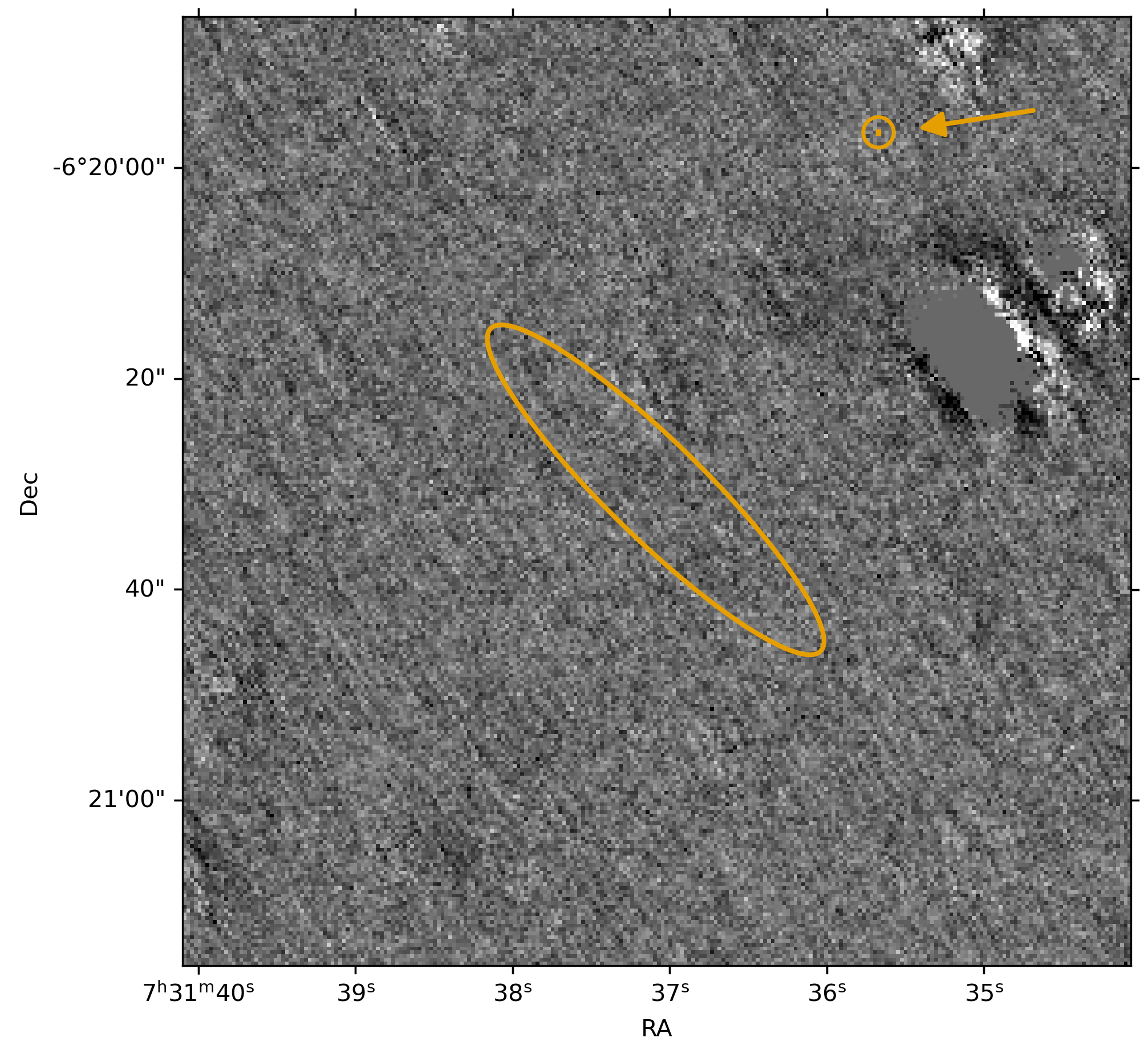}
    \caption{The LDT composite, star-subtracted image on 2020 October 16 centered at the nominal position of C/2019 Y4-B. The 3$\sigma$ uncertainty ellipse based on JPL orbit solution \#9 is shown.}
    \label{fig:detection}
\end{figure}

Figure~\ref{fig:detection} shows the stacked, star-subtracted image obtained on 2020 October 16, overlaid with the uncertainty ellipse of JPL orbit \#9 (Figure~\ref{fig:detection}), which shows no plausible source. Based on the photometric calibration results and a $3''$-radius aperture, we derived a $3\sigma$ limit of $r=25.5$. To constrain the fragment's absolute magnitude, we used the following relation:  
\begin{equation}
H=r-5\log{\left(\frac{d_{BS}d_{BO}}{d_0^2}\right)}-0.035\alpha,
\end{equation}

\noindent where $r$ is the apparent magnitude, $d_{BS}$ is the distance between the fragment and the Sun in au, $d_{BO}$ is the distance between the observer and the object in au, $d_0$ is the average distance between the Sun and the Earth (1 au), and $\alpha$ is the phase angle in degrees. Using the appropriate values, we derived an absolute magnitude of $H>20.5$. This implies that an inactive fragment B was at $r>24.4$~mag at the earlier August 20 observation ($r_\mathrm{h}=1.84$~au), well below the image limit, and was therefore not detectable.

To constrain the fragment's effective size using its absolute magnitude, we modeled the fragment as a smooth sphere with a uniform surface albedo. Under these assumptions, the following expression approximates the fragment's diameter: 

\begin{equation}
D\approx \frac{1329}{\sqrt{p}}10^{-0.2H},
\end{equation}

\noindent where $D$ is the diameter of the fragment in km, $p=0.04$ is its albedo \citep[assumed to be the average of comets, cf.][]{lamy2004sizes}, and $H$ is the absolute magnitude. Using this expression, we derive an upper limit on the diameter of fragment B of $D\lesssim 0.5$ km. 

\subsubsection{ZTF}
 
The LDT detection implied that fragment B was too faint for ZTF to detect ($r>25.5$ versus a general limiting magnitude of $r\sim21$). However, ZTF provided a wider coverage in time. Therefore, we visually inspected the ZTF nightly stacks in a similar manner to fragment A. We did not find fragment B in this data set, suggesting that it did not experience a sudden increase in brightness during this period. 

\section{Discussion}

Our constraint on the diameter of C/2019 Y4-B's remnant --- $D\lesssim 0.5$~km --- is significantly smaller than the pre-perihelion observations made six months ago ($D\sim1.2$~km, Paper I), indicating that fragment B had undergone further major disintegration. However, we cannot conclude from the available data whether any sizable fragment(s) still exist. Observed disintegration events have produced long-lasting fragments as small as $D\sim0.3$~km \citep[e.g. 289P/Blanpain;][]{2006AJ....131.2327J}, which is smaller than our detection limit.

The non-detection of C/2019 Y4-A is of little surprise owing to the shallower survey images and the fact that fragment A had likely fully disrupted even before perihelion. The pre-perihelion HST observations showed that fragment A exhibited a much higher rate of change of size distribution, losing $70\%$ of its mass within a 3-day period \citep{Ye2021}, implying that the fragment may have completely disintegrated shortly after the HST observations. Independent analysis performed by \citet{sekanina2023fragmentation} also argued that C/2019 Y4's initial surge in brightness was primarily driven by fragment A's rapid disintegration and consequent increase in cross-sectional area of released dust rather than by the relatively inactive fragment B.

Nevertheless, the uncertain fate of C/2019 Y4 raises an intriguing question: how many presumably disrupted comets have really completely disrupted, and could any of them have actually survived with a reduced, inactive nucleus? As a cursory exploration, we review the long-period comets considered to have disintegrated or fragmented as listed in \citet{2022AJ....164..158J}. The author did not make a clear distinction between fragmented or disintegrated comets, categorizing the comets merely based on their reported morphological behaviors for simplicity of their discussion. C/2019 Y4 was categorized as ``fragmented,'' presumably based on the reports that the comet first split into multiple components.

As shown in Table~\ref{tab:dead_comets}, 4 out of 7 of the cases involve presumably disintegrated comets that have not been reportedly confirmed, making up a significant fraction of the sample. With the exception of C/2019 Y4, which is studied in this work, three out of these four cases are because of the lack of follow-up confirmation, likely due to poor observability immediately after disintegration, although none of these are more difficult than the case of C/2019 Y4: the observability generally improved 2--3 months after the solar conjunction. For C/2019 Y4, a deep search right after the solar conjunction (such as immediately following the initial shallower search in early August of 2020) could have conclusively determined the state of the remnant. Similarly, dedicated deep searches would be helpful in closing cases like the other three comets and would provide insights into comet disintegration dynamics. 

\begin{deluxetable*}{ccl}
\tablecaption{Review of presumably disrupted comets listed in \citet{2022AJ....164..158J}. \label{tab:dead_comets}}
\tablehead{
    \colhead{Comet} & \colhead{Fate} & \colhead{Comment}
}
\startdata
C/2001 A2 (LINEAR) & Primary component survived & Multiple transient fragments \citep{2002ApJ...572..679S} \\
C/2010 X1 (Elenin) & Complete disruption & CFHT non-detection of $r>40$~m fragments \citep{2015AJ....149..133L} \\
C/2012 S1 (ISON) & Presumed disruption & No published post-fragmentation search results \\
C/2017 E4 (Lovejoy) & Presumed disruption & No published post-fragmentation search results \\
C/2019 Y4 (ATLAS) & Presumed disruption & This work \\
C/2020 F8 (SWAN) & Presumed disruption & No published post-fragmentation search results \\
C/2021 A1 (Leonard) & Complete disruption & HST non-detection of $r>60$~m fragments \citep{Jewitt_2023} \\
\enddata
\end{deluxetable*}

\section{Summary}

We report our attempt to recover disrupted comet C/2019 Y4 (ATLAS) using LDT and data from ZTF in late 2020, 3--5 months after the comet's significant disruption and the last published observation. We constrained any possible remnant of the primary component, C/2019 Y4-B, to a diameter of $D<0.5$ km, significantly smaller than the pre-perihelion measurement of $D\sim1.2$~km, suggesting that the fragment has experienced further major disintegration since its last observation before the perihelion. We also searched for fragments C/2019 Y4-A in shallower ZTF images that covered a wider range of time in which the comet receded from $r_\mathrm{h}=2.5$ to 2.9~au without any detection, placing an upper-bound of $D\lesssim2$ km for fragment A.

The uncertain fate of C/2019 Y4 highlights the challenge in understanding the state of presumably disrupted comets. Our review of six other presumably disrupted long-period comets reported in the recent decade shows that half of these cases have not been negatively confirmed, rendering their disruption presumptions inconclusive. Dedicated deep follow-up observations of future cases like these may clarify the fate of these disrupted comets and further our understanding of comet disintegration.


\begin{acknowledgments}
We thank an anonymous referee for comments, as well as Victoria Girgis, Ishara Nisley, Sydney Perez, Jason Sanborn, and LaLaina Shumar for assisting the Lowell Discovery Telescope (LDT) observations. We also thank Matthew Knight for helpful discussion and Davide Farnocchia for checking the orbit fit of our recovery detection. The LDT observations were acquired by the University of Maryland observing team consisting Gerbs Bauer, Adeline Gicquel-Brodtke, Tony Farnham, Lori Feaga, Michael Kelley, Jessica Sunshine, and Quanzhi Ye. SCQ and QY are supported by NASA/ESA Hubble Space Telescope grant \#16089 and NASA program 80NSSC22K0772. MSPK is supported by NASA program 80NSSC20K0673. These results made use of LDT at Lowell Observatory. Lowell is a private, non-profit institution dedicated to astrophysical research and public appreciation of astronomy and operates the LDT in partnership with Boston University, the University of Maryland, the University of Toledo, Northern Arizona University and Yale University. ZTF observations were obtained with the Samuel Oschin 48-inch Telescope at the Palomar Observatory as part of the Zwicky Transient Facility project. ZTF is supported by the National Science Foundation under Grant No. AST-1440341 and a collaboration including Caltech, IPAC, the Weizmann Institute for Science, the Oskar Klein Center at Stockholm University, the University of Maryland, the University of Washington, Deutsches Elektronen-Synchrotron and Humboldt University, Los Alamos National Laboratories, the TANGO Consortium of Taiwan, the University of Wisconsin at Milwaukee, and Lawrence Berkeley National Laboratories. Operations are conducted by COO, IPAC, and UW. This work is also supported by Boston University's Undergraduate Research Opportunities Program. 
\end{acknowledgments}

\vspace{5mm}
\facilities{LDT, PO:1.2m}

\software{{\tt ccdproc} \citep{Craig2015}, {\tt dct-redux} \citep{2024zndo..13946957K}, {\tt HOTPANTS \citep{2015ascl.soft04004B}, {\tt ZChecker} \citep{Kelley2019}}
          }

\bibliography{sample631}{}
\bibliographystyle{aasjournal}



\end{CJK*}
\end{document}